\PassOptionsToPackage{unicode}{hyperref}
\PassOptionsToPackage{hyphens}{url}
\documentclass[
]{article}
\usepackage{lmodern}
\usepackage{amssymb,amsmath}
\usepackage{ifxetex,ifluatex}
\ifnum 0\ifxetex 1\fi\ifluatex 1\fi=0 
  \usepackage[T1]{fontenc}
  \usepackage[utf8]{inputenc}
  \usepackage{textcomp} 
\else 
  \usepackage{unicode-math}
  \defaultfontfeatures{Scale=MatchLowercase}
  \defaultfontfeatures[\rmfamily]{Ligatures=TeX,Scale=1}
\fi
\IfFileExists{upquote.sty}{\usepackage{upquote}}{}
\IfFileExists{microtype.sty}{
  \usepackage[]{microtype}
  \UseMicrotypeSet[protrusion]{basicmath} 
}{}
\makeatletter
\@ifundefined{KOMAClassName}{
  \IfFileExists{parskip.sty}{%
    \usepackage{parskip}
  }{
    \setlength{\parindent}{0pt}
    \setlength{\parskip}{6pt plus 2pt minus 1pt}}
}{
  \KOMAoptions{parskip=half}}
\makeatother
\usepackage{xcolor}
\IfFileExists{xurl.sty}{\usepackage{xurl}}{} 
\IfFileExists{bookmark.sty}{\usepackage{bookmark}}{\usepackage{hyperref}}
\hypersetup{
  pdftitle={Brazilian Obstetric Observatory},
  pdfauthor={Agatha Rodrigues\^{}\{1,2\}, Lucas Lacerda\^{}1, Rossana Francisco\^{}2},
  hidelinks,
  pdfcreator={LaTeX via pandoc}}
\usepackage[margin=1in]{geometry}
\usepackage{color}
\usepackage{fancyvrb}

\DefineVerbatimEnvironment{Highlighting}{Verbatim}{commandchars=\\\{\}}
\usepackage{framed}
\definecolor{shadecolor}{RGB}{248,248,248}
\newenvironment{Shaded}{\begin{snugshade}}{\end{snugshade}}

\newcommand{\CommentTok}[1]{\textcolor[rgb]{0.56,0.35,0.01}{\textit{#1}}}

\newcommand{\ControlFlowTok}[1]{\textcolor[rgb]{0.13,0.29,0.53}{\textbf{#1}}}
\newcommand{\DataTypeTok}[1]{\textcolor[rgb]{0.13,0.29,0.53}{#1}}
\newcommand{\DecValTok}[1]{\textcolor[rgb]{0.00,0.00,0.81}{#1}}

\newcommand{\KeywordTok}[1]{\textcolor[rgb]{0.13,0.29,0.53}{\textbf{#1}}}
\newcommand{\NormalTok}[1]{#1}
\newcommand{\OperatorTok}[1]{\textcolor[rgb]{0.81,0.36,0.00}{\textbf{#1}}}
\newcommand{\OtherTok}[1]{\textcolor[rgb]{0.56,0.35,0.01}{#1}}

\newcommand{\StringTok}[1]{\textcolor[rgb]{0.31,0.60,0.02}{#1}}

\usepackage{longtable,booktabs}
\usepackage{etoolbox}
\makeatletter
\patchcmd\longtable{\par}{\if@noskipsec\mbox{}\fi\par}{}{}
\makeatother
\IfFileExists{footnotehyper.sty}{\usepackage{footnotehyper}}{\usepackage{footnote}}
\makesavenoteenv{longtable}
\usepackage{graphicx,grffile}
\makeatletter
\def\maxwidth{\ifdim\Gin@nat@width>\linewidth\linewidth\else\Gin@nat@width\fi}
\def\maxheight{\ifdim\Gin@nat@height>\textheight\textheight\else\Gin@nat@height\fi}
\makeatother
\setkeys{Gin}{width=\maxwidth,height=\maxheight,keepaspectratio}
\makeatletter
\def\fps@figure{htbp}
\makeatother
\title{Brazilian Obstetric Observatory}
\author{Agatha Rodrigues\(^{1,2}\), Lucas Lacerda\(^1\), Rossana Francisco\(^2\)}
\date{\(^1\)Department of Statistics of Federal University of Espírito Santo;
\(^2\)Department of Obstetrics and Gynecology of University of São Paulo}

\begin{document}
\maketitle

\hypertarget{abstract}{%
\section{Abstract}\label{abstract}}

Covid-19 is responsible for high mortality in all countries, with the
maternal population it is no different. Countries with a high rate of
maternal mortality have deficiencies in the health care of pregnant
women and women who have recently given birth, which will certainly be
enhanced in a situation of overload in the health system, as occurred in
this pandemic. Understanding the impact of the pandemic on maternal
health is essential to discuss public policies and assist in solutions
to future crises. With that in mind, we present the Brazilian Obstetric
Observatory COVID-19 (OOBr COVID-19). OOBr COVID-19 is a dynamic panel
with analyzes of the cases of pregnant and postpartum women with Severe
Acute Respiratory Syndrome (SARI) during the pandemic due to the new
coronavirus. In this article, we present data loading, case selections,
and processing of the variables for the analyzes available in OOBr
COVID-19.

\hypertarget{introduction}{%
\section{1. Introduction}\label{introduction}}

Covid-19 (disease caused by SARS-CoV-2) has been responsible for high
mortality in all countries. In November 2020, {[}1{]} pointed out that
pregnant women would have a higher risk of hospitalization in intensive
care units, orotracheal intubation and death than non-pregnant women.

Covid-19 presented many clinical manifestations and has shown
inequalities among countries especially with regard to access to
healthcare systems. The difference in the mortality rate of pregnant and
postpartum women in the world by COVID-19 reflects the differences among
countries' maternal death rates observed before the pandemic caused by
COVID-19. Countries with a high rate of maternal death have deficiencies
in healthcare for pregnant women and women who have recently given
birth, which will certainly be enhanced in a situation of overload to
the healthcare system, as occurred in this pandemic.

The Brazilian Obstetric Observatory COVID-19 (OOBr COVID-19, in
Portuguese: Observatório Obstétrico Brasileiro COVID-19) is a dynamic
panel with analyzes of the cases of pregnant and postpartum women with
Severe Acute Respiratory Syndrome (SARI) during the pandemic due to the
new coronavirus. The OOBr COVID-19 aims to give visibility to the data
of this specific public and to offer tools for analysis and reasoning
for health care policies for pregnant women and women who have recently
given birth.

There are considered the records of reports in the SIVEP Gripe database
(Influenza Epidemiological Surveillance Information System), a
nationwide surveillance database used to monitor SARI in Brazil. The
database is made available by the Ministry of Health of Brazil and
updated weekly on the website
\url{https://opendatasus.saude.gov.br/dataset}.

Notification is mandatory for Influenza Syndrome (characterized by at
least two of the following signs and symptoms: fever, even if referred,
chills, sore throat, headache, cough, runny nose, olfactory or taste
disorders) and who has dyspnea/respiratory discomfort or persistent
pressure in the chest or O2 saturation less than 95\% in room air or
bluish color of the lips or face. Asymptomatic individuals with
laboratory confirmation by molecular biology or immunological
examination for COVID-19 infection are also reported. For notifications
in Sivep-Gripe, hospitalized cases in both public and private hospitals
and all deaths due to severe acute respiratory infections regardless of
hospitalization must be considered.

OOBr COVID-19 can be accessed at
\url{https://observatorioobstetrico.shinyapps.io/covid_gesta_puerp_br}.
The analyzed period comprised data from epidemiological weeks 8 to 53 of
2020 (12/29/2019 - 01/02/2021) and from epidemiological weeks 1 to until
the last available update of 2021. The OOBr COVID-19 is updated weekly
depending on the updates made available by the Ministry of Health on the
website \url{https://opendatasus.saude.gov.br/dataset} {[}2,3{]}.

In this article, we will describe the selections, filters, and data
transformations to achieve the information available in OOBr COVID-19.
Section 2 describes the methods used to select the cases and process the
variables for OOBr COVID-19 and the results after loading the data and
processing the variables are presented in Section 3. Finally, the final
remarks are presented in Section 4.

\hypertarget{methods}{%
\section{2. Methods}\label{methods}}

The database from 2020 is downloaded weekly in
\url{https://opendatasus.saude.gov.br/dataset/bd-srag-2020} and the
database from 2021 is downloaded weekly in
\url{https://opendatasus.saude.gov.br/dataset/bd-srag-2021}.

The data are analyzed using the R program, version 4.0.3
(\url{https://www.r-project.org}) and the OOBr COVID-19 is available on
a Shiny dashboard (\url{https://www.shinyapps.io}).

The two databases are merged and data are filtered from the 8th
epidemiological week of symptoms (when the first confirmed case of
COVID-19 was found in the database) to until the last available update
of 2021. All data on female cases aged 10 to 55 years old, with
information on whether pregnant women (first, second or third
gestational period or with ignored gestational age) or in the puerperium
period were included. The flowchart is shown in Figure 1.

\begin{figure}
\centering
\includegraphics{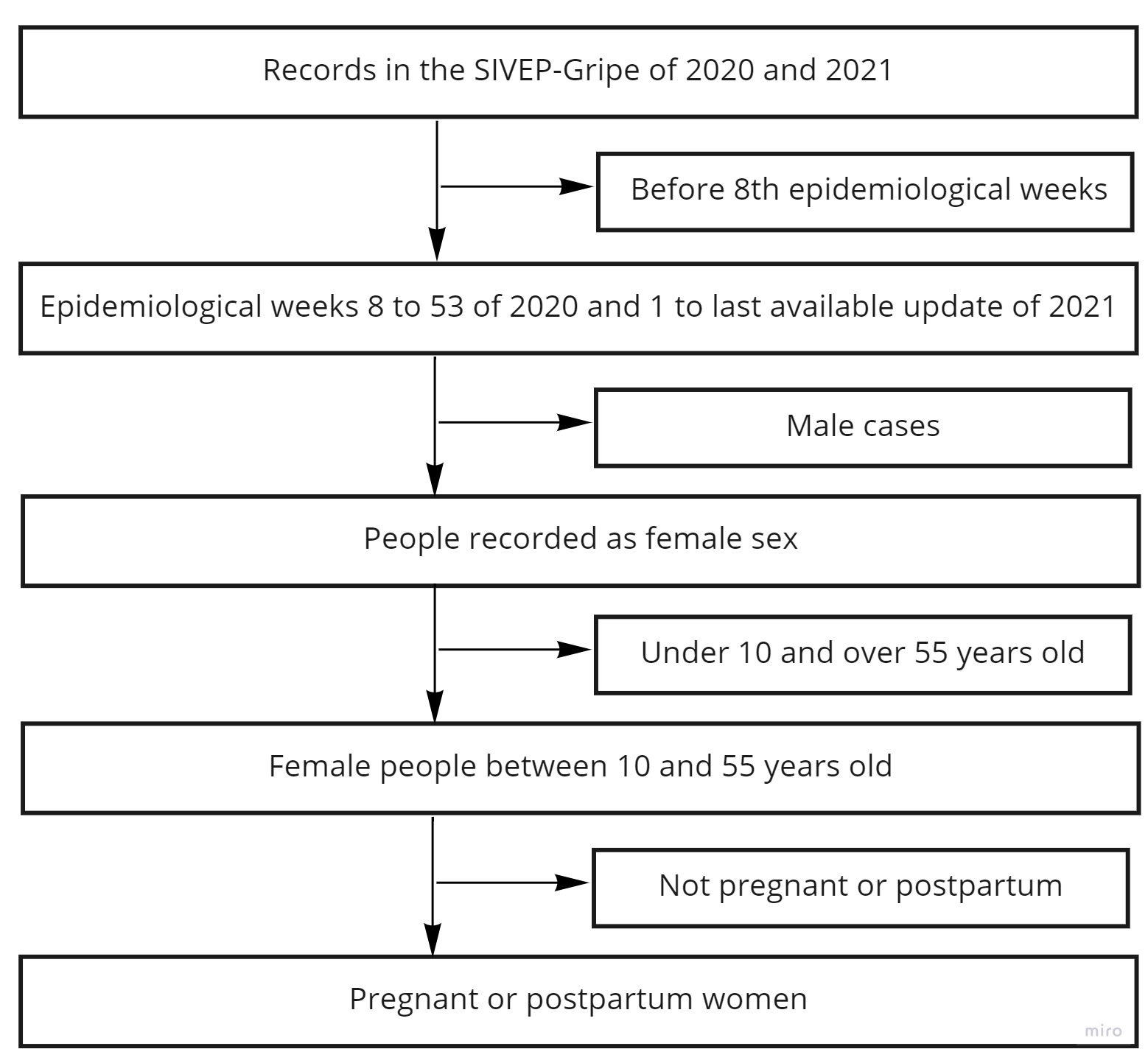}
\caption{Flowchart of case selection}
\end{figure}

The variables analyzed and available in OOBr COVID-19 are: age, race,
education, state of Brazil of residence, region of Brazil of residence,
age range, obstetric status, change of municipality for assistance,
residence area, SARI diagnosis, laboratory (etiological) diagnosis, flu
syndrome that progresses to SARI, type of antiviral, previous
vaccination for influenza, hospital-acquired infection, travel history,
contact with swine, signs and symptoms (fever, cough, sore throat,
dyspnoea, respiratory distress, O2 saturation less than 95\%, diarrhea,
vomiting, abdominal pain, fatigue, loss of smell or taste), risk
factors/comorbidities (cardiovascular disease, kidney disease,
neurological disease, hematological disease, liver disease, diabetes,
asthma, pneumopathy, obesity and immunosuppression), hospitalization,
admission to the ICU (Intensive Care Unit), use of ventilatory support
(invasive and non-invasive) and evolution of the case (cure or death).

\hypertarget{results}{%
\section{3. Results}\label{results}}

The analyzes that result in observatory
\url{https://observatorioobstetrico.shinyapps.io/covid_gesta_puerp_br}
are described in this section. At first, the R packages used are
presented, the data are loaded, the selections and filters are made and,
finally, the characterization variables, symptoms, comorbidities, and
outcome variables are processed.

\hypertarget{database-load-and-r-packages-used}{%
\subsection{3.1 Database load and R packages
used}\label{database-load-and-r-packages-used}}

The R packages used for filtering and data processing are presented in
this subsection.

\begin{Shaded}
\begin{Highlighting}[]
\CommentTok{#R packages used}
\NormalTok{loadlibrary <-}\StringTok{ }\ControlFlowTok{function}\NormalTok{(x) \{}
  \ControlFlowTok{if}\NormalTok{ (}\OperatorTok{!}\KeywordTok{require}\NormalTok{(x, }\DataTypeTok{character.only =} \OtherTok{TRUE}\NormalTok{)) \{}
    \KeywordTok{install.packages}\NormalTok{(x, }\DataTypeTok{dependencies =}\NormalTok{ T)}
    \ControlFlowTok{if}\NormalTok{ (}\OperatorTok{!}\KeywordTok{require}\NormalTok{(x, }\DataTypeTok{character.only =} \OtherTok{TRUE}\NormalTok{))}
      \KeywordTok{stop}\NormalTok{(}\StringTok{"Package not found"}\NormalTok{)}
\NormalTok{  \}}
\NormalTok{\}}

\NormalTok{packages <-}
\StringTok{  }\KeywordTok{c}\NormalTok{(}
    \StringTok{"readr"}\NormalTok{,}
    \StringTok{"readxl"}\NormalTok{,}
    \StringTok{"janitor"}\NormalTok{,}
    \StringTok{"dplyr"}\NormalTok{,}
    \StringTok{"forcats"}\NormalTok{,}
    \StringTok{"stringr"}\NormalTok{,}
    \StringTok{"lubridate"}\NormalTok{,}
    \StringTok{"summarytools"}\NormalTok{,}
    \StringTok{"magrittr"}\NormalTok{,}
    \StringTok{"questionr"}\NormalTok{, }
    \StringTok{"knitr"}
\NormalTok{  )}
\KeywordTok{lapply}\NormalTok{(packages, loadlibrary)}
\end{Highlighting}
\end{Shaded}

The 2020 and 2021 databases are loaded and they are also merged. Below
are the databases updated on April 26, 2021, the last update available
at the time of writing this article.

\begin{Shaded}
\begin{Highlighting}[]
\CommentTok{######### Importing databases}
\CommentTok{#2021}
\NormalTok{dados_}\DecValTok{2021}\NormalTok{ <-}\StringTok{ }\KeywordTok{read_delim}\NormalTok{(}
  \StringTok{"INFLUD21-26-04-2021.csv"}\NormalTok{,}
  \StringTok{";"}\NormalTok{,}
  \DataTypeTok{escape_double =} \OtherTok{FALSE}\NormalTok{,}
  \DataTypeTok{locale =} \KeywordTok{locale}\NormalTok{(}\DataTypeTok{encoding =} \StringTok{"ISO-8859-2"}\NormalTok{),}
  \DataTypeTok{trim_ws =} \OtherTok{TRUE}
\NormalTok{)}

\CommentTok{#2020}
\NormalTok{dados_}\DecValTok{2020}\NormalTok{ <-}\StringTok{ }\KeywordTok{read_delim}\NormalTok{(}
  \StringTok{"INFLUD-26-04-2021.csv"}\NormalTok{,}
  \StringTok{";"}\NormalTok{,}
  \DataTypeTok{escape_double =} \OtherTok{FALSE}\NormalTok{,}
  \DataTypeTok{locale =} \KeywordTok{locale}\NormalTok{(}\DataTypeTok{encoding =} \StringTok{"ISO-8859-2"}\NormalTok{),}
  \DataTypeTok{trim_ws =} \OtherTok{TRUE}
\NormalTok{)}

\CommentTok{####### Merging 2020 and 2021 databases }
\NormalTok{dados1 <-}\StringTok{ }\KeywordTok{rbind}\NormalTok{(dados_}\DecValTok{2020}\NormalTok{, dados_}\DecValTok{2021}\NormalTok{)}
\end{Highlighting}
\end{Shaded}

\hypertarget{selecting-cases-and-data-processing}{%
\subsection{3.2 Selecting cases and data
processing}\label{selecting-cases-and-data-processing}}

We will filter only the cases from the 8th epidemiological week of 2020
(first confirmed case of COVID-19) until the current epidemiological
week of 2021.

\begin{Shaded}
\begin{Highlighting}[]
\CommentTok{#### Current epidemiological week }
\NormalTok{sem <-}\StringTok{ }\DecValTok{16} 

\CommentTok{#### Create year variable (ano)}
\NormalTok{dados1 <-}\StringTok{  }\NormalTok{dados1 }\OperatorTok{%>%}
\StringTok{  }\NormalTok{dplyr}\OperatorTok{::}\KeywordTok{mutate}\NormalTok{(}
    \DataTypeTok{dt_sint =} \KeywordTok{as.Date}\NormalTok{(DT_SIN_PRI, }\DataTypeTok{format =} \StringTok{"%d/%m/%Y"}\NormalTok{),}
    \DataTypeTok{ano =}\NormalTok{ lubridate}\OperatorTok{::}\KeywordTok{year}\NormalTok{(dt_sint),}
\NormalTok{  )}

\CommentTok{#### Case filtering from the 8th epidemiological week of 2020}
\NormalTok{dados2 <-}\StringTok{ }\NormalTok{dados1 }\OperatorTok{%>%}\StringTok{ }
\StringTok{  }\KeywordTok{filter}\NormalTok{((ano}\OperatorTok{==}\DecValTok{2020} \OperatorTok{&}\StringTok{ }\NormalTok{SEM_PRI }\OperatorTok{>=}\DecValTok{8}\NormalTok{) }\OperatorTok{|}\StringTok{ }\NormalTok{ano }\OperatorTok{==}\DecValTok{2021}\NormalTok{)}
\end{Highlighting}
\end{Shaded}

The table in the following presents the distribution of cases by year
and by epidemiological week.

\begin{Shaded}
\begin{Highlighting}[]
\CommentTok{#### Cross table of epidemiological year and week  }
\KeywordTok{ctable}\NormalTok{(dados2}\OperatorTok{$}\NormalTok{SEM_PRI, dados2}\OperatorTok{$}\NormalTok{ano, }\DataTypeTok{prop=}\StringTok{"n"}\NormalTok{)}
\end{Highlighting}
\end{Shaded}

\begin{verbatim}
## Cross-Tabulation  
## SEM_PRI * ano  
## Data Frame: dados2  
## 
## --------- ----- --------- -------- ---------
##             ano      2020     2021     Total
##   SEM_PRI                                   
##         1               0    35122     35122
##         2               0    33809     33809
##         3               0    31044     31044
##         4               0    28985     28985
##         5               0    34670     34670
##         6               0    37745     37745
##         7               0    47328     47328
##         8             923    50091     51014
##         9            1164    68575     69739
##        10            1980    68057     70037
##        11            5136    65397     70533
##        12           12826    50657     63483
##        13           14974    45334     60308
##        14           16289    37006     53295
##        15           19583    19824     39407
##        16           24872     4445     29317
##        17           30828       19     30847
##        18           34864        0     34864
##        19           34566        0     34566
##        20           37172        0     37172
##        21           33824        0     33824
##        22           31256        0     31256
##        23           35643        0     35643
##        24           34162        0     34162
##        25           36694        0     36694
##        26           32968        0     32968
##        27           37444        0     37444
##        28           37041        0     37041
##        29           34440        0     34440
##        30           33704        0     33704
##        31           32184        0     32184
##        32           30028        0     30028
##        33           31070        0     31070
##        34           28271        0     28271
##        35           26337        0     26337
##        36           26459        0     26459
##        37           24047        0     24047
##        38           22222        0     22222
##        39           21579        0     21579
##        40           22451        0     22451
##        41           21026        0     21026
##        42           18999        0     18999
##        43           19464        0     19464
##        44           18719        0     18719
##        45           23302        0     23302
##        46           25802        0     25802
##        47           29289        0     29289
##        48           29162        0     29162
##        49           32940        0     32940
##        50           30540        0     30540
##        51           28405        0     28405
##        52           30304        0     30304
##        53           21559    12514     34073
##     Total         1176512   670622   1847134
## --------- ----- --------- -------- ---------
\end{verbatim}

Note that there are 12514 cases in 2021 in week 53. These are cases from
the first two days of 2021, which are still part of of the last
epidemiological week of 2020
(\url{http://portalsinan.saude.gov.br/calendario-epidemiologico?layout=edit\&id=168}).
However, these cases belong to the 53rd week of 2020 and we corrected in
the following:

\begin{Shaded}
\begin{Highlighting}[]
\CommentTok{#### Correcting year variable (ano) from 53rd epidemiological week}
\NormalTok{dados2 <-}\StringTok{ }\NormalTok{dados2 }\OperatorTok{%>%}\StringTok{ }
\StringTok{  }\KeywordTok{mutate}\NormalTok{(}\DataTypeTok{ano =} \KeywordTok{ifelse}\NormalTok{(ano }\OperatorTok{==}\DecValTok{2021} \OperatorTok{&}\StringTok{ }\NormalTok{SEM_PRI }\OperatorTok{==}\DecValTok{53}\NormalTok{, }\DecValTok{2020}\NormalTok{, ano)) }\OperatorTok{%>%}\StringTok{ }
\StringTok{  }\KeywordTok{filter}\NormalTok{(ano}\OperatorTok{==}\DecValTok{2020} \OperatorTok{|}\StringTok{ }\NormalTok{(ano }\OperatorTok{==}\StringTok{ }\DecValTok{2021} \OperatorTok{&}\StringTok{ }\NormalTok{SEM_PRI }\OperatorTok{<=}\StringTok{ }\NormalTok{sem)) }
\end{Highlighting}
\end{Shaded}

The distribution of epidemiological week by year of the pandemic after
correction is presented in the following.

\begin{Shaded}
\begin{Highlighting}[]
\CommentTok{#### Cross table of epidemiological year and week }
\KeywordTok{ctable}\NormalTok{(dados2}\OperatorTok{$}\NormalTok{SEM_PRI, dados2}\OperatorTok{$}\NormalTok{ano, }\DataTypeTok{prop=}\StringTok{"n"}\NormalTok{)}
\end{Highlighting}
\end{Shaded}

\begin{verbatim}
## Cross-Tabulation  
## SEM_PRI * ano  
## Data Frame: dados2  
## 
## --------- ----- --------- -------- ---------
##             ano      2020     2021     Total
##   SEM_PRI                                   
##         1               0    35122     35122
##         2               0    33809     33809
##         3               0    31044     31044
##         4               0    28985     28985
##         5               0    34670     34670
##         6               0    37745     37745
##         7               0    47328     47328
##         8             923    50091     51014
##         9            1164    68575     69739
##        10            1980    68057     70037
##        11            5136    65397     70533
##        12           12826    50657     63483
##        13           14974    45334     60308
##        14           16289    37006     53295
##        15           19583    19824     39407
##        16           24872     4445     29317
##        17           30828        0     30828
##        18           34864        0     34864
##        19           34566        0     34566
##        20           37172        0     37172
##        21           33824        0     33824
##        22           31256        0     31256
##        23           35643        0     35643
##        24           34162        0     34162
##        25           36694        0     36694
##        26           32968        0     32968
##        27           37444        0     37444
##        28           37041        0     37041
##        29           34440        0     34440
##        30           33704        0     33704
##        31           32184        0     32184
##        32           30028        0     30028
##        33           31070        0     31070
##        34           28271        0     28271
##        35           26337        0     26337
##        36           26459        0     26459
##        37           24047        0     24047
##        38           22222        0     22222
##        39           21579        0     21579
##        40           22451        0     22451
##        41           21026        0     21026
##        42           18999        0     18999
##        43           19464        0     19464
##        44           18719        0     18719
##        45           23302        0     23302
##        46           25802        0     25802
##        47           29289        0     29289
##        48           29162        0     29162
##        49           32940        0     32940
##        50           30540        0     30540
##        51           28405        0     28405
##        52           30304        0     30304
##        53           34073        0     34073
##     Total         1189026   658089   1847115
## --------- ----- --------- -------- ---------
\end{verbatim}

The next step is to identify pregnant women. For this, we will analyze
the variable \texttt{CS\_GESTANT}. This variable assumes the values:
1-1st trimester; 2-2nd trimester; 3-3rd trimester; 4-Ignored Gestational
Age; 5-No; 6-Does not apply; 9-Ignored.

\begin{Shaded}
\begin{Highlighting}[]
\CommentTok{##### Frequency table for gestational information}
\NormalTok{questionr}\OperatorTok{::}\KeywordTok{freq}\NormalTok{(}
\NormalTok{  dados2}\OperatorTok{$}\NormalTok{CS_GESTANT,}
  \DataTypeTok{cum =} \OtherTok{FALSE}\NormalTok{,}
  \DataTypeTok{total =} \OtherTok{TRUE}\NormalTok{,}
  \DataTypeTok{na.last =} \OtherTok{FALSE}\NormalTok{,}
  \DataTypeTok{valid =} \OtherTok{FALSE}
\NormalTok{) }\OperatorTok{%>%}
\StringTok{  }\KeywordTok{kable}\NormalTok{(}\DataTypeTok{caption =} \StringTok{"Frequency table for variable}
\StringTok{         about pregnancy"}\NormalTok{, }\DataTypeTok{digits =} \DecValTok{2}\NormalTok{) }
\end{Highlighting}
\end{Shaded}

\begin{longtable}[]{@{}lrr@{}}
\caption{Frequency table for variable about pregnancy}\tabularnewline
\toprule
& n & \%\tabularnewline
\midrule
\endfirsthead
\toprule
& n & \%\tabularnewline
\midrule
\endhead
0 & 371 & 0.0\tabularnewline
1 & 1938 & 0.1\tabularnewline
2 & 4475 & 0.2\tabularnewline
3 & 9593 & 0.5\tabularnewline
4 & 978 & 0.1\tabularnewline
5 & 574545 & 31.1\tabularnewline
6 & 1163859 & 63.0\tabularnewline
9 & 91356 & 4.9\tabularnewline
Total & 1847115 & 100.0\tabularnewline
\bottomrule
\end{longtable}

There are 371 cases with \texttt{CS\_GESTANT=0}, where category 0 has no
code in the database dictionary.

The next step is to check if there is any inconsistency when analyzing
this variable together with sex (\texttt{CS\_SEXO}), with categories
F-female, M-male and I-ignored.

\begin{Shaded}
\begin{Highlighting}[]
\CommentTok{#### Cross table of gestation and sex}
\KeywordTok{ctable}\NormalTok{(dados2}\OperatorTok{$}\NormalTok{CS_GESTANT, dados2}\OperatorTok{$}\NormalTok{CS_SEXO, }\DataTypeTok{prop=}\StringTok{"n"}\NormalTok{)}
\end{Highlighting}
\end{Shaded}

\begin{verbatim}
## Cross-Tabulation  
## CS_GESTANT * CS_SEXO  
## Data Frame: dados2  
## 
## ------------ --------- -------- ----- -------- ---------
##                CS_SEXO        F     I        M     Total
##   CS_GESTANT                                            
##            0                115   177       79       371
##            1               1938     0        0      1938
##            2               4474     1        0      4475
##            3               9592     1        0      9593
##            4                977     1        0       978
##            5             573354    51     1140    574545
##            6             165319   279   998261   1163859
##            9              91108    94      154     91356
##        Total             846877   604   999634   1847115
## ------------ --------- -------- ----- -------- ---------
\end{verbatim}

There are 0 cases of \texttt{CS\_SEXO=M} with
\texttt{CS\_GESTANT=1,2,3\ ou\ 4}, hopefully.

The puerperium indicator variable is \texttt{PUERPERA}, with categories
1-yes, 2-no and 9-Ignored.

\begin{Shaded}
\begin{Highlighting}[]
\CommentTok{#Frequency table for puerperium}
\NormalTok{questionr}\OperatorTok{::}\KeywordTok{freq}\NormalTok{(}
\NormalTok{  dados2}\OperatorTok{$}\NormalTok{PUERPERA,}
  \DataTypeTok{cum =} \OtherTok{FALSE}\NormalTok{,}
  \DataTypeTok{total =} \OtherTok{TRUE}\NormalTok{,}
  \DataTypeTok{na.last =} \OtherTok{FALSE}\NormalTok{,}
  \DataTypeTok{valid =} \OtherTok{FALSE}
\NormalTok{) }\OperatorTok{%>%}
\StringTok{  }\KeywordTok{kable}\NormalTok{(}\DataTypeTok{caption =} \StringTok{"Frequency table for puerperium"}\NormalTok{, }\DataTypeTok{digits =} \DecValTok{2}\NormalTok{)}
\end{Highlighting}
\end{Shaded}

\begin{longtable}[]{@{}lrr@{}}
\caption{Frequency table for puerperium}\tabularnewline
\toprule
& n & \%\tabularnewline
\midrule
\endfirsthead
\toprule
& n & \%\tabularnewline
\midrule
\endhead
1 & 6648 & 0.4\tabularnewline
2 & 682995 & 37.0\tabularnewline
9 & 18082 & 1.0\tabularnewline
NA & 1139390 & 61.7\tabularnewline
Total & 1847115 & 100.0\tabularnewline
\bottomrule
\end{longtable}

The next step is to check if there is any inconsistency when analyzing
this variable together with sex (\texttt{CS\_SEXO}), with categories
F-female, M-male and I-ignored.

\begin{Shaded}
\begin{Highlighting}[]
\CommentTok{#### Cross table of puerperium and sex}
\KeywordTok{ctable}\NormalTok{(dados2}\OperatorTok{$}\NormalTok{PUERPERA, dados2}\OperatorTok{$}\NormalTok{CS_SEXO, }\DataTypeTok{prop=}\StringTok{"n"}\NormalTok{)}
\end{Highlighting}
\end{Shaded}

\begin{verbatim}
## Cross-Tabulation  
## PUERPERA * CS_SEXO  
## Data Frame: dados2  
## 
## ---------- --------- -------- ----- -------- ---------
##              CS_SEXO        F     I        M     Total
##   PUERPERA                                            
##          1               6647     1        0      6648
##          2             327159   174   355662    682995
##          9               8345    10     9727     18082
##       <NA>             504726   419   634245   1139390
##      Total             846877   604   999634   1847115
## ---------- --------- -------- ----- -------- ---------
\end{verbatim}

There are 0 cases of \texttt{CS\_SEXO=M} with \texttt{PUERPERA\ =\ 1},
that is, puerperium and male sex cases, hopefully.

The next selection is to consider only female people and aged over 10
and under or equal to 55 years.

\begin{Shaded}
\begin{Highlighting}[]
\CommentTok{#### Filtering only female cases}
\NormalTok{dados3 <-}\StringTok{ }\NormalTok{dados2 }\OperatorTok{%>%}\StringTok{ }
\StringTok{  }\KeywordTok{filter}\NormalTok{(CS_SEXO }\OperatorTok{==}\StringTok{ "F"}\NormalTok{)}

\CommentTok{#### Filtering of cases aged 55 years or less}
\NormalTok{dados4 <-}\StringTok{ }\NormalTok{dados3 }\OperatorTok{%>%}\StringTok{ }
\StringTok{  }\KeywordTok{filter}\NormalTok{(NU_IDADE_N }\OperatorTok{>}\StringTok{ }\DecValTok{9} \OperatorTok{&}\StringTok{ }\NormalTok{NU_IDADE_N }\OperatorTok{<=}\StringTok{ }\DecValTok{55}\NormalTok{)}
\end{Highlighting}
\end{Shaded}

Now we are going to create the variable of gestational trimester or
puerperium. Note that for puerperium (\texttt{puerp}), nonpregnant or
ignored cases with \texttt{PUERPERA\ =\ 1} are considered.

\begin{Shaded}
\begin{Highlighting}[]
\CommentTok{#### Creation of the classi_gesta_puerp variable for the gestational or postpartum period}
\NormalTok{dados4 <-}\StringTok{ }\NormalTok{dados4 }\OperatorTok{%>%}
\StringTok{  }\KeywordTok{mutate}\NormalTok{(}
    \DataTypeTok{classi_gesta_puerp =} \KeywordTok{case_when}\NormalTok{(}
\NormalTok{      CS_GESTANT }\OperatorTok{==}\StringTok{ }\DecValTok{1}  \OperatorTok{~}\StringTok{ "1tri"}\NormalTok{, }\CommentTok{# 1st trimester}
\NormalTok{      CS_GESTANT }\OperatorTok{==}\StringTok{ }\DecValTok{2}  \OperatorTok{~}\StringTok{ "2tri"}\NormalTok{, }\CommentTok{# 2nd trimester}
\NormalTok{      CS_GESTANT }\OperatorTok{==}\StringTok{ }\DecValTok{3}  \OperatorTok{~}\StringTok{ "3tri"}\NormalTok{, }\CommentTok{# 3rd trimester}
\NormalTok{      CS_GESTANT }\OperatorTok{==}\StringTok{ }\DecValTok{4}  \OperatorTok{~}\StringTok{ "IG_ig"}\NormalTok{, }\CommentTok{# Ignored gestational age}
\NormalTok{      CS_GESTANT }\OperatorTok{==}\StringTok{ }\DecValTok{5} \OperatorTok{&}
\StringTok{        }\NormalTok{PUERPERA }\OperatorTok{==}\StringTok{ }\DecValTok{1} \OperatorTok{~}\StringTok{ "puerp"}\NormalTok{, }\CommentTok{# puerperal woman}
\NormalTok{      CS_GESTANT }\OperatorTok{==}\StringTok{ }\DecValTok{9} \OperatorTok{&}\StringTok{ }\NormalTok{PUERPERA }\OperatorTok{==}\StringTok{ }\DecValTok{1} \OperatorTok{~}\StringTok{ "puerp"}\NormalTok{, }\CommentTok{# puerperal woman}
      \OtherTok{TRUE} \OperatorTok{~}\StringTok{ "não"} \CommentTok{# 'não' means not pregnant}
\NormalTok{    )}
\NormalTok{  )}
\end{Highlighting}
\end{Shaded}

The last filtering consists of selecting the cases of pregnant or
postpartum women.

\begin{Shaded}
\begin{Highlighting}[]
\CommentTok{### Selection only of pregnant or postpartum cases}
\NormalTok{dados5 <-}\StringTok{ }\NormalTok{dados4 }\OperatorTok{%>%}
\StringTok{  }\KeywordTok{filter}\NormalTok{(classi_gesta_puerp }\OperatorTok{!=}\StringTok{ "não"}\NormalTok{)}
\end{Highlighting}
\end{Shaded}

\begin{Shaded}
\begin{Highlighting}[]
\CommentTok{### Frequency table for gestational group}
\NormalTok{questionr}\OperatorTok{::}\KeywordTok{freq}\NormalTok{(}
\NormalTok{  dados5}\OperatorTok{$}\NormalTok{classi_gesta_puerp,}
  \DataTypeTok{cum =} \OtherTok{FALSE}\NormalTok{,}
  \DataTypeTok{total =} \OtherTok{TRUE}\NormalTok{,}
  \DataTypeTok{na.last =} \OtherTok{FALSE}\NormalTok{,}
  \DataTypeTok{valid =} \OtherTok{FALSE}
\NormalTok{) }\OperatorTok{%>%}
\StringTok{  }\KeywordTok{kable}\NormalTok{(}\DataTypeTok{caption =} \StringTok{"Frequency table for gestational trimester or postpartum variable"}\NormalTok{,}
        \DataTypeTok{digits =} \DecValTok{2}\NormalTok{)}
\end{Highlighting}
\end{Shaded}

\begin{longtable}[]{@{}lrr@{}}
\caption{Frequency table for gestational trimester or postpartum
variable}\tabularnewline
\toprule
& n & \%\tabularnewline
\midrule
\endfirsthead
\toprule
& n & \%\tabularnewline
\midrule
\endhead
1tri & 1914 & 8.9\tabularnewline
2tri & 4407 & 20.5\tabularnewline
3tri & 9575 & 44.6\tabularnewline
IG\_ig & 917 & 4.3\tabularnewline
puerp & 4661 & 21.7\tabularnewline
Total & 21474 & 100.0\tabularnewline
\bottomrule
\end{longtable}

In the following, we deal with other variables considered at the OOBr
Covid-19.

\hypertarget{variables-processing}{%
\subsection{3.3 Variables processing}\label{variables-processing}}

The variable that indicates the SARI diagnosis is \texttt{CLASSI\_FIN},
with categories: 1-SARI by influenza, 2-SARI by another respiratory
virus, 3-SARI by another etiologic agent, 4-SARI not specified, 5-SARI
by COVID-19.

\begin{Shaded}
\begin{Highlighting}[]
\CommentTok{#frequency table for SARI diagnosis}
\NormalTok{questionr}\OperatorTok{::}\KeywordTok{freq}\NormalTok{(}
\NormalTok{  dados5}\OperatorTok{$}\NormalTok{CLASSI_FIN,}
  \DataTypeTok{cum =} \OtherTok{FALSE}\NormalTok{,}
  \DataTypeTok{total =} \OtherTok{TRUE}\NormalTok{,}
  \DataTypeTok{na.last =} \OtherTok{FALSE}\NormalTok{,}
  \DataTypeTok{valid =} \OtherTok{FALSE}
\NormalTok{) }\OperatorTok{%>%}
\StringTok{  }\KeywordTok{kable}\NormalTok{(}\DataTypeTok{caption =} \StringTok{"Frequency table for SARI diagnosis"}\NormalTok{, }\DataTypeTok{digits =} \DecValTok{2}\NormalTok{)}
\end{Highlighting}
\end{Shaded}

\begin{longtable}[]{@{}lrr@{}}
\caption{Frequency table for SARI diagnosis}\tabularnewline
\toprule
& n & \%\tabularnewline
\midrule
\endfirsthead
\toprule
& n & \%\tabularnewline
\midrule
\endhead
1 & 74 & 0.3\tabularnewline
2 & 105 & 0.5\tabularnewline
3 & 61 & 0.3\tabularnewline
4 & 8104 & 37.7\tabularnewline
5 & 10818 & 50.4\tabularnewline
NA & 2312 & 10.8\tabularnewline
Total & 21474 & 100.0\tabularnewline
\bottomrule
\end{longtable}

The variable that identify the diagnostic type of COVID-19 is
\texttt{classi\_covid}, with categories: pcr (RT-PCR), antigenio
(antigen), sorologia (serology) and outro (other). This variable only
has valid categories for cases confirmed by SARI by COVID-19
(\texttt{CLASSI\_FIN=5}).

\begin{Shaded}
\begin{Highlighting}[]
\CommentTok{#Case diagnosed by RT-PCR}
\NormalTok{dados5 <-}\StringTok{ }\NormalTok{dados5 }\OperatorTok{%>%}
\StringTok{  }\KeywordTok{mutate}\NormalTok{(}\DataTypeTok{pcr_SN =} \KeywordTok{case_when}\NormalTok{(}
\NormalTok{    (PCR_SARS2 }\OperatorTok{==}\StringTok{ }\DecValTok{1}\NormalTok{) }\OperatorTok{|}
\StringTok{      }\NormalTok{(}
        \KeywordTok{str_detect}\NormalTok{(DS_PCR_OUT, }\StringTok{"SARS|COVID|COV|CORONA|CIVID"}\NormalTok{) }
\NormalTok{      ) }\OperatorTok{~}\StringTok{ "sim"}\NormalTok{,}
    \OtherTok{TRUE} \OperatorTok{~}\StringTok{ "não"}
\NormalTok{  ))}

\CommentTok{#Identify if diagnosed by serology}
\NormalTok{dados5}\OperatorTok{$}\NormalTok{res_igg <-}
\StringTok{  }\KeywordTok{ifelse}\NormalTok{(}\KeywordTok{is.na}\NormalTok{(dados5}\OperatorTok{$}\NormalTok{RES_IGG) }\OperatorTok{==}\StringTok{ }\OtherTok{TRUE}\NormalTok{, }\DecValTok{0}\NormalTok{, dados5}\OperatorTok{$}\NormalTok{RES_IGG)}

\NormalTok{dados5}\OperatorTok{$}\NormalTok{res_igm <-}
\StringTok{  }\KeywordTok{ifelse}\NormalTok{(}\KeywordTok{is.na}\NormalTok{(dados5}\OperatorTok{$}\NormalTok{RES_IGM) }\OperatorTok{==}\StringTok{ }\OtherTok{TRUE}\NormalTok{, }\DecValTok{0}\NormalTok{, dados5}\OperatorTok{$}\NormalTok{RES_IGM)}

\NormalTok{dados5}\OperatorTok{$}\NormalTok{res_iga <-}
\StringTok{  }\KeywordTok{ifelse}\NormalTok{(}\KeywordTok{is.na}\NormalTok{(dados5}\OperatorTok{$}\NormalTok{RES_IGA) }\OperatorTok{==}\StringTok{ }\OtherTok{TRUE}\NormalTok{, }\DecValTok{0}\NormalTok{, dados5}\OperatorTok{$}\NormalTok{RES_IGA)}

\NormalTok{dados5}\OperatorTok{$}\NormalTok{sorologia_SN <-}
\StringTok{  }\KeywordTok{ifelse}\NormalTok{(dados5}\OperatorTok{$}\NormalTok{res_igg }\OperatorTok{==}\StringTok{ }\DecValTok{1} \OperatorTok{|}
\StringTok{           }\NormalTok{dados5}\OperatorTok{$}\NormalTok{res_igm }\OperatorTok{==}\StringTok{ }\DecValTok{1} \OperatorTok{|}\StringTok{ }\NormalTok{dados5}\OperatorTok{$}\NormalTok{res_iga }\OperatorTok{==}\StringTok{ }\DecValTok{1}\NormalTok{,}
         \StringTok{"sim"}\NormalTok{,}
         \StringTok{"não"}\NormalTok{)}

\CommentTok{#Identify if diagnosed by antigen}
\NormalTok{dados5 <-}\StringTok{ }\NormalTok{dados5 }\OperatorTok{%>%}
\StringTok{  }\KeywordTok{mutate}\NormalTok{(}\DataTypeTok{antigeno_SN =} \KeywordTok{case_when}\NormalTok{(}
\NormalTok{    (AN_SARS2 }\OperatorTok{==}\StringTok{ }\DecValTok{1}\NormalTok{) }\OperatorTok{|}\StringTok{ }\CommentTok{#positivo}
\StringTok{      }\NormalTok{(}
        \KeywordTok{str_detect}\NormalTok{(DS_AN_OUT, }\StringTok{"SARS|COVID|COV|CORONA|CONA"}\NormalTok{) }
\NormalTok{      )  }\OperatorTok{~}\StringTok{ "sim"}\NormalTok{,}
    \OtherTok{TRUE} \OperatorTok{~}\StringTok{ "não"}
\NormalTok{  ))}

\CommentTok{#Creation of the covid-19 classification variable}
\NormalTok{dados5 <-}\StringTok{ }\NormalTok{dados5 }\OperatorTok{%>%}
\StringTok{  }\KeywordTok{mutate}\NormalTok{(}
    \DataTypeTok{classi_covid =} \KeywordTok{case_when}\NormalTok{(}
\NormalTok{      CLASSI_FIN }\OperatorTok{==}\StringTok{ }\DecValTok{5} \OperatorTok{&}\StringTok{ }\NormalTok{pcr_SN }\OperatorTok{==}\StringTok{ "sim"}  \OperatorTok{~}\StringTok{ "pcr"}\NormalTok{,}
\NormalTok{      CLASSI_FIN }\OperatorTok{==}\StringTok{ }\DecValTok{5} \OperatorTok{&}\StringTok{ }\NormalTok{pcr_SN }\OperatorTok{==}\StringTok{ "não"} \OperatorTok{&}
\StringTok{        }\NormalTok{antigeno_SN }\OperatorTok{==}\StringTok{ "sim"} \OperatorTok{~}\StringTok{ "antigenio"}\NormalTok{,}
\NormalTok{      CLASSI_FIN }\OperatorTok{==}\StringTok{ }\DecValTok{5} \OperatorTok{&}\StringTok{ }\NormalTok{sorologia_SN }\OperatorTok{==}\StringTok{ "sim"} \OperatorTok{&}
\StringTok{        }\NormalTok{antigeno_SN }\OperatorTok{==}\StringTok{ "não"} \OperatorTok{&}
\StringTok{        }\NormalTok{pcr_SN }\OperatorTok{==}\StringTok{ "não"} \OperatorTok{~}\StringTok{ "sorologia"}\NormalTok{,}
\NormalTok{      CLASSI_FIN }\OperatorTok{!=}\StringTok{ }\DecValTok{5} \OperatorTok{~}\StringTok{ "não"}\NormalTok{, }\CommentTok{#is not another etiologic agent or unspecified}
      \OtherTok{TRUE} \OperatorTok{~}\StringTok{ "outro"}
\NormalTok{    )}
\NormalTok{  )}
\end{Highlighting}
\end{Shaded}

\begin{Shaded}
\begin{Highlighting}[]
\CommentTok{#frequency table for COVID-19 diagnostic type}
\NormalTok{questionr}\OperatorTok{::}\KeywordTok{freq}\NormalTok{(}
\NormalTok{  dados5}\OperatorTok{$}\NormalTok{classi_covid,}
  \DataTypeTok{cum =} \OtherTok{FALSE}\NormalTok{,}
  \DataTypeTok{total =} \OtherTok{TRUE}\NormalTok{,}
  \DataTypeTok{na.last =} \OtherTok{FALSE}\NormalTok{,}
  \DataTypeTok{valid =} \OtherTok{FALSE}
\NormalTok{) }\OperatorTok{%>%}
\StringTok{  }\KeywordTok{kable}\NormalTok{(}\DataTypeTok{caption =} \StringTok{"Frequency table for the COVID-19 diagnostic type"}\NormalTok{, }\DataTypeTok{digits =} \DecValTok{2}\NormalTok{)}
\end{Highlighting}
\end{Shaded}

\begin{longtable}[]{@{}lrr@{}}
\caption{Frequency table for the COVID-19 diagnostic
type}\tabularnewline
\toprule
& n & \%\tabularnewline
\midrule
\endfirsthead
\toprule
& n & \%\tabularnewline
\midrule
\endhead
antigenio & 827 & 3.9\tabularnewline
não & 8344 & 38.9\tabularnewline
outro & 4354 & 20.3\tabularnewline
pcr & 6908 & 32.2\tabularnewline
sorologia & 1041 & 4.8\tabularnewline
Total & 21474 & 100.0\tabularnewline
\bottomrule
\end{longtable}

The variable that indicates the state of Brazil is \texttt{SG\_UF}. The
variable that indicates the region of Brazil (North, Northeast, Central,
Southeast and South) is \texttt{region}, created in the following.

\begin{Shaded}
\begin{Highlighting}[]
\CommentTok{#Creation of the region variable}
\NormalTok{regions <-}\StringTok{ }\ControlFlowTok{function}\NormalTok{(state) \{}
\NormalTok{  southeast <-}\StringTok{ }\KeywordTok{c}\NormalTok{(}\StringTok{"SP"}\NormalTok{, }\StringTok{"RJ"}\NormalTok{, }\StringTok{"ES"}\NormalTok{, }\StringTok{"MG"}\NormalTok{)}
\NormalTok{  south <-}\StringTok{ }\KeywordTok{c}\NormalTok{(}\StringTok{"PR"}\NormalTok{, }\StringTok{"SC"}\NormalTok{, }\StringTok{"RS"}\NormalTok{)}
\NormalTok{  central <-}\StringTok{ }\KeywordTok{c}\NormalTok{(}\StringTok{"GO"}\NormalTok{, }\StringTok{"MT"}\NormalTok{, }\StringTok{"MS"}\NormalTok{, }\StringTok{"DF"}\NormalTok{)}
\NormalTok{  northeast <-}
\StringTok{    }\KeywordTok{c}\NormalTok{(}\StringTok{"AL"}\NormalTok{, }\StringTok{"BA"}\NormalTok{, }\StringTok{"CE"}\NormalTok{, }\StringTok{"MA"}\NormalTok{, }\StringTok{"PB"}\NormalTok{, }\StringTok{"PE"}\NormalTok{, }\StringTok{"PI"}\NormalTok{, }\StringTok{"RN"}\NormalTok{, }\StringTok{"SE"}\NormalTok{)}
\NormalTok{  north <-}\StringTok{ }\KeywordTok{c}\NormalTok{(}\StringTok{"AC"}\NormalTok{, }\StringTok{"AP"}\NormalTok{, }\StringTok{"AM"}\NormalTok{, }\StringTok{"PA"}\NormalTok{, }\StringTok{"RO"}\NormalTok{, }\StringTok{"RR"}\NormalTok{, }\StringTok{"TO"}\NormalTok{)}
\NormalTok{  out <-}
\StringTok{    }\KeywordTok{ifelse}\NormalTok{(}\KeywordTok{any}\NormalTok{(state }\OperatorTok{==}\StringTok{ }\NormalTok{southeast),}
           \StringTok{"southeast"}\NormalTok{,}
           \KeywordTok{ifelse}\NormalTok{(}\KeywordTok{any}\NormalTok{(state }\OperatorTok{==}\StringTok{ }\NormalTok{south),}
                  \StringTok{"south"}\NormalTok{,}
                  \KeywordTok{ifelse}\NormalTok{(}
                    \KeywordTok{any}\NormalTok{(state }\OperatorTok{==}\StringTok{ }\NormalTok{central),}
                    \StringTok{"central"}\NormalTok{,}
                    \KeywordTok{ifelse}\NormalTok{(}\KeywordTok{any}\NormalTok{(state }\OperatorTok{==}\StringTok{ }\NormalTok{northeast),}
                           \StringTok{"northeast"}\NormalTok{, }\StringTok{"north"}\NormalTok{)}
\NormalTok{                  )))}
  \KeywordTok{return}\NormalTok{(out)}
\NormalTok{\}}

\NormalTok{dados5}\OperatorTok{$}\NormalTok{region <-}\StringTok{ }\KeywordTok{sapply}\NormalTok{(dados5}\OperatorTok{$}\NormalTok{SG_UF, regions)}
\NormalTok{dados5}\OperatorTok{$}\NormalTok{region <-}
\StringTok{  }\KeywordTok{ifelse}\NormalTok{(}\KeywordTok{is.na}\NormalTok{(dados5}\OperatorTok{$}\NormalTok{region) }\OperatorTok{==}\StringTok{ }\OtherTok{TRUE}\NormalTok{, }\DecValTok{0}\NormalTok{, dados5}\OperatorTok{$}\NormalTok{region)}
\end{Highlighting}
\end{Shaded}

\begin{Shaded}
\begin{Highlighting}[]
\CommentTok{#Frequency table for region}
\NormalTok{questionr}\OperatorTok{::}\KeywordTok{freq}\NormalTok{(}
\NormalTok{  dados5}\OperatorTok{$}\NormalTok{region,}
  \DataTypeTok{cum =} \OtherTok{FALSE}\NormalTok{,}
  \DataTypeTok{total =} \OtherTok{TRUE}\NormalTok{,}
  \DataTypeTok{na.last =} \OtherTok{FALSE}\NormalTok{,}
  \DataTypeTok{valid =} \OtherTok{FALSE}
\NormalTok{) }\OperatorTok{%>%}
\StringTok{  }\KeywordTok{kable}\NormalTok{(}\DataTypeTok{caption =} \StringTok{"Frequency table for the region of Brazil"}\NormalTok{, }\DataTypeTok{digits =} \DecValTok{2}\NormalTok{)}
\end{Highlighting}
\end{Shaded}

\begin{longtable}[]{@{}lrr@{}}
\caption{Frequency table for the region of Brazil}\tabularnewline
\toprule
& n & \%\tabularnewline
\midrule
\endfirsthead
\toprule
& n & \%\tabularnewline
\midrule
\endhead
0 & 4 & 0.0\tabularnewline
central & 2384 & 11.1\tabularnewline
north & 2363 & 11.0\tabularnewline
northeast & 5394 & 25.1\tabularnewline
south & 2671 & 12.4\tabularnewline
southeast & 8658 & 40.3\tabularnewline
Total & 21474 & 100.0\tabularnewline
\bottomrule
\end{longtable}

Note that there are 4 cases without information for the region of the
country (encoded as 0).

The processing of the characterization variables is presented in the
following.

\begin{Shaded}
\begin{Highlighting}[]
\CommentTok{#Race}
\NormalTok{dados5 <-}\StringTok{  }\NormalTok{dados5 }\OperatorTok{%>%}
\StringTok{  }\KeywordTok{mutate}\NormalTok{(}
    \DataTypeTok{raca =} \KeywordTok{case_when}\NormalTok{(}
\NormalTok{      CS_RACA }\OperatorTok{==}\StringTok{ }\DecValTok{1} \OperatorTok{~}\StringTok{ "branca"}\NormalTok{, }\CommentTok{#white}
\NormalTok{      CS_RACA }\OperatorTok{==}\StringTok{ }\DecValTok{2} \OperatorTok{~}\StringTok{ "preta"}\NormalTok{,   }\CommentTok{#black}
\NormalTok{      CS_RACA }\OperatorTok{==}\StringTok{ }\DecValTok{3} \OperatorTok{~}\StringTok{ "amarela"}\NormalTok{, }\CommentTok{#yellow}
\NormalTok{      CS_RACA }\OperatorTok{==}\StringTok{ }\DecValTok{4} \OperatorTok{~}\StringTok{ "parda"}\NormalTok{,   }\CommentTok{#brown}
\NormalTok{      CS_RACA }\OperatorTok{==}\StringTok{ }\DecValTok{5} \OperatorTok{~}\StringTok{ "indigena"}\NormalTok{, }\CommentTok{#indigenous}
      \OtherTok{TRUE} \OperatorTok{~}\StringTok{ }\OtherTok{NA_character_}
\NormalTok{    )}
\NormalTok{  )}

\CommentTok{#Education}
\NormalTok{dados5 <-}\StringTok{  }\NormalTok{dados5 }\OperatorTok{%>%}
\StringTok{  }\KeywordTok{mutate}\NormalTok{(}
    \DataTypeTok{escol =} \KeywordTok{case_when}\NormalTok{(}
\NormalTok{      CS_ESCOL_N }\OperatorTok{==}\StringTok{ }\DecValTok{0} \OperatorTok{~}\StringTok{ "sem escol"}\NormalTok{, }\CommentTok{#no school}
\NormalTok{      CS_ESCOL_N }\OperatorTok{==}\StringTok{ }\DecValTok{1} \OperatorTok{~}\StringTok{ "fund1"}\NormalTok{,     }\CommentTok{#1st elementary school}
\NormalTok{      CS_ESCOL_N }\OperatorTok{==}\StringTok{ }\DecValTok{2} \OperatorTok{~}\StringTok{ "fund2"}\NormalTok{,   }\CommentTok{#2nd elementary school}
\NormalTok{      CS_ESCOL_N }\OperatorTok{==}\StringTok{ }\DecValTok{3} \OperatorTok{~}\StringTok{ "medio"}\NormalTok{,   }\CommentTok{#high school}
\NormalTok{      CS_ESCOL_N }\OperatorTok{==}\StringTok{ }\DecValTok{4} \OperatorTok{~}\StringTok{ "superior"}\NormalTok{, }\CommentTok{#university education}
      \OtherTok{TRUE} \OperatorTok{~}\StringTok{ }\OtherTok{NA_character_}
\NormalTok{    )}
\NormalTok{  )}

\CommentTok{#Age range}
\NormalTok{dados5 <-}\StringTok{  }\NormalTok{dados5 }\OperatorTok{%>%}
\StringTok{  }\KeywordTok{mutate}\NormalTok{(}
    \DataTypeTok{faixa_et =} \KeywordTok{case_when}\NormalTok{(}
\NormalTok{      NU_IDADE_N }\OperatorTok{<=}\StringTok{ }\DecValTok{19} \OperatorTok{~}\StringTok{ "<20"}\NormalTok{,}
\NormalTok{      NU_IDADE_N }\OperatorTok{>=}\StringTok{ }\DecValTok{20}
      \OperatorTok{&}\StringTok{ }\NormalTok{NU_IDADE_N }\OperatorTok{<=}\StringTok{ }\DecValTok{34} \OperatorTok{~}\StringTok{ "20-34"}\NormalTok{,}
\NormalTok{      NU_IDADE_N }\OperatorTok{>=}\StringTok{ }\DecValTok{35} \OperatorTok{~}\StringTok{ ">=35"}\NormalTok{,}
      \OtherTok{TRUE} \OperatorTok{~}\StringTok{ }\OtherTok{NA_character_}
\NormalTok{    )}
\NormalTok{  )}
\NormalTok{dados5}\OperatorTok{$}\NormalTok{faixa_et <-}
\StringTok{  }\KeywordTok{factor}\NormalTok{(dados5}\OperatorTok{$}\NormalTok{faixa_et, }\DataTypeTok{levels =} \KeywordTok{c}\NormalTok{(}\StringTok{"<20"}\NormalTok{, }\StringTok{"20-34"}\NormalTok{, }\StringTok{">=35"}\NormalTok{))}

\CommentTok{#Hospitalization}
\NormalTok{dados5 <-}\StringTok{  }\NormalTok{dados5 }\OperatorTok{%>%}
\StringTok{  }\KeywordTok{mutate}\NormalTok{(}\DataTypeTok{hospital =} \KeywordTok{case_when}\NormalTok{(HOSPITAL }\OperatorTok{==}\StringTok{ }\DecValTok{1} \OperatorTok{~}\StringTok{ "sim"}\NormalTok{, }\CommentTok{#yes}
\NormalTok{                              HOSPITAL }\OperatorTok{==}\StringTok{ }\DecValTok{2} \OperatorTok{~}\StringTok{ "não"}\NormalTok{, }\CommentTok{#no}
                              \OtherTok{TRUE} \OperatorTok{~}\StringTok{ }\OtherTok{NA_character_}\NormalTok{))}

\CommentTok{#Travel history}
\NormalTok{dados5 <-}\StringTok{  }\NormalTok{dados5 }\OperatorTok{%>%}
\StringTok{  }\KeywordTok{mutate}\NormalTok{(}\DataTypeTok{hist_viagem =} \KeywordTok{case_when}\NormalTok{(HISTO_VGM }\OperatorTok{==}\StringTok{ }\DecValTok{1} \OperatorTok{~}\StringTok{ "sim"}\NormalTok{, }\CommentTok{#yes}
\NormalTok{                                 HISTO_VGM }\OperatorTok{==}\StringTok{ }\DecValTok{2} \OperatorTok{~}\StringTok{ "não"}\NormalTok{, }\CommentTok{#no}
                                 \OtherTok{TRUE} \OperatorTok{~}\StringTok{ }\OtherTok{NA_character_}\NormalTok{))}

\CommentTok{#Influenza syndrome evolved to SARI}
\NormalTok{dados5 <-}\StringTok{  }\NormalTok{dados5 }\OperatorTok{%>%}
\StringTok{  }\KeywordTok{mutate}\NormalTok{(}\DataTypeTok{sg_para_srag =} \KeywordTok{case_when}\NormalTok{(SURTO_SG }\OperatorTok{==}\StringTok{ }\DecValTok{1} \OperatorTok{~}\StringTok{ "sim"}\NormalTok{, }\CommentTok{#yes}
\NormalTok{                                  SURTO_SG }\OperatorTok{==}\StringTok{ }\DecValTok{2} \OperatorTok{~}\StringTok{ "não"}\NormalTok{, }\CommentTok{#no}
                                  \OtherTok{TRUE} \OperatorTok{~}\StringTok{ }\OtherTok{NA_character_}\NormalTok{))}

\CommentTok{#Hospital acquired infection}
\NormalTok{dados5 <-}\StringTok{  }\NormalTok{dados5 }\OperatorTok{%>%}
\StringTok{  }\KeywordTok{mutate}\NormalTok{(}\DataTypeTok{inf_inter =} \KeywordTok{case_when}\NormalTok{(NOSOCOMIAL }\OperatorTok{==}\StringTok{ }\DecValTok{1} \OperatorTok{~}\StringTok{ "sim"}\NormalTok{, }\CommentTok{#yes}
\NormalTok{                               NOSOCOMIAL }\OperatorTok{==}\StringTok{ }\DecValTok{2} \OperatorTok{~}\StringTok{ "não"}\NormalTok{, }\CommentTok{#no}
                               \OtherTok{TRUE} \OperatorTok{~}\StringTok{ }\OtherTok{NA_character_}\NormalTok{))}

\CommentTok{#Contact with poultry or swine}
\NormalTok{dados5 <-}\StringTok{  }\NormalTok{dados5 }\OperatorTok{%>%}
\StringTok{  }\KeywordTok{mutate}\NormalTok{(}\DataTypeTok{cont_ave_suino =} \KeywordTok{case_when}\NormalTok{(AVE_SUINO }\OperatorTok{==}\StringTok{ }\DecValTok{1} \OperatorTok{~}\StringTok{ "sim"}\NormalTok{, }\CommentTok{#yes}
\NormalTok{                                    AVE_SUINO }\OperatorTok{==}\StringTok{ }\DecValTok{2} \OperatorTok{~}\StringTok{ "não"}\NormalTok{, }\CommentTok{#no}
                                    \OtherTok{TRUE} \OperatorTok{~}\StringTok{ }\OtherTok{NA_character_}\NormalTok{))}

\CommentTok{#Influenza vaccine}
\NormalTok{dados5 <-}\StringTok{  }\NormalTok{dados5 }\OperatorTok{%>%}
\StringTok{  }\KeywordTok{mutate}\NormalTok{(}\DataTypeTok{vacina =} \KeywordTok{case_when}\NormalTok{(VACINA }\OperatorTok{==}\StringTok{ }\DecValTok{1} \OperatorTok{~}\StringTok{ "sim"}\NormalTok{, }\CommentTok{#yes}
\NormalTok{                            VACINA }\OperatorTok{==}\StringTok{ }\DecValTok{2} \OperatorTok{~}\StringTok{ "não"}\NormalTok{, }\CommentTok{#no}
                            \OtherTok{TRUE} \OperatorTok{~}\StringTok{ }\OtherTok{NA_character_}\NormalTok{))}

\CommentTok{#Antiviral}
\NormalTok{dados5 <-}\StringTok{  }\NormalTok{dados5 }\OperatorTok{%>%}
\StringTok{  }\KeywordTok{mutate}\NormalTok{(}
    \DataTypeTok{antiviral =} \KeywordTok{case_when}\NormalTok{(}
\NormalTok{      ANTIVIRAL }\OperatorTok{==}\StringTok{ }\DecValTok{1} \OperatorTok{~}\StringTok{ "Oseltamivir"}\NormalTok{,}
\NormalTok{      ANTIVIRAL }\OperatorTok{==}\StringTok{ }\DecValTok{2} \OperatorTok{~}\StringTok{ "Zanamivir"}\NormalTok{,}
      \OtherTok{TRUE} \OperatorTok{~}\StringTok{ }\OtherTok{NA_character_}\NormalTok{  ))}

\CommentTok{#Residence zone}
\NormalTok{dados5 <-}\StringTok{  }\NormalTok{dados5 }\OperatorTok{%>%}
\StringTok{  }\KeywordTok{mutate}\NormalTok{(}\DataTypeTok{zona =} \KeywordTok{case_when}\NormalTok{(CS_ZONA }\OperatorTok{==}\StringTok{ }\DecValTok{1} \OperatorTok{~}\StringTok{ "urbana"}\NormalTok{, }\CommentTok{#urban}
\NormalTok{                          CS_ZONA }\OperatorTok{==}\StringTok{ }\DecValTok{2} \OperatorTok{~}\StringTok{ "rural"}\NormalTok{,  }\CommentTok{#rural}
\NormalTok{                          CS_ZONA }\OperatorTok{==}\StringTok{ }\DecValTok{3} \OperatorTok{~}\StringTok{ "periurbana"}\NormalTok{, }\CommentTok{#periurban}
                                  \OtherTok{TRUE} \OperatorTok{~}\StringTok{ }\OtherTok{NA_character_}\NormalTok{))}

\CommentTok{#If change of municipality for care}
\NormalTok{dados5 <-}\StringTok{ }\NormalTok{dados5 }\OperatorTok{%>%}\StringTok{ }
\StringTok{  }\KeywordTok{mutate}\NormalTok{(}\DataTypeTok{mudou_muni =} \KeywordTok{case_when}\NormalTok{(CO_MUN_RES}\OperatorTok{==}\NormalTok{CO_MU_INTE }\OperatorTok{&}\StringTok{ }\OperatorTok{!}\KeywordTok{is.na}\NormalTok{(CO_MU_INTE) }\OperatorTok{&}\StringTok{ }
\StringTok{                                  }\OperatorTok{!}\KeywordTok{is.na}\NormalTok{(CO_MUN_RES) }\OperatorTok{~}\StringTok{ "não"}\NormalTok{, }\CommentTok{#no}
\NormalTok{                                CO_MUN_RES}\OperatorTok{!=}\NormalTok{CO_MU_INTE }\OperatorTok{&}\StringTok{ }\OperatorTok{!}\KeywordTok{is.na}\NormalTok{(CO_MU_INTE) }\OperatorTok{&}\StringTok{ }
\StringTok{                                  }\OperatorTok{!}\KeywordTok{is.na}\NormalTok{(CO_MUN_RES) }\OperatorTok{~}\StringTok{ "sim"}\NormalTok{, }\CommentTok{#yes}
                                \OtherTok{TRUE} \OperatorTok{~}\StringTok{ }\OtherTok{NA_character_}
\NormalTok{                                )}
\NormalTok{  )}
\end{Highlighting}
\end{Shaded}

The processing of symptom variables is presented below.

\begin{Shaded}
\begin{Highlighting}[]
\CommentTok{#Fever}
\NormalTok{dados5 <-}\StringTok{  }\NormalTok{dados5 }\OperatorTok{%>%}
\StringTok{  }\KeywordTok{mutate}\NormalTok{(}\DataTypeTok{febre =} \KeywordTok{case_when}\NormalTok{(FEBRE }\OperatorTok{==}\StringTok{ }\DecValTok{1} \OperatorTok{~}\StringTok{ "sim"}\NormalTok{, }\CommentTok{#yes}
\NormalTok{                           FEBRE }\OperatorTok{==}\StringTok{ }\DecValTok{2} \OperatorTok{~}\StringTok{ "não"}\NormalTok{, }\CommentTok{#no}
                           \OtherTok{TRUE} \OperatorTok{~}\StringTok{ }\OtherTok{NA_character_}\NormalTok{))}

\CommentTok{#Cough}
\NormalTok{dados5 <-}\StringTok{  }\NormalTok{dados5 }\OperatorTok{%>%}
\StringTok{  }\KeywordTok{mutate}\NormalTok{(}\DataTypeTok{tosse =} \KeywordTok{case_when}\NormalTok{(TOSSE }\OperatorTok{==}\StringTok{ }\DecValTok{1} \OperatorTok{~}\StringTok{ "sim"}\NormalTok{, }\CommentTok{#yes}
\NormalTok{                           TOSSE }\OperatorTok{==}\StringTok{ }\DecValTok{2} \OperatorTok{~}\StringTok{ "não"}\NormalTok{, }\CommentTok{#no}
                           \OtherTok{TRUE} \OperatorTok{~}\StringTok{ }\OtherTok{NA_character_}\NormalTok{))}

\CommentTok{#Sore throat}
\NormalTok{dados5 <-}\StringTok{  }\NormalTok{dados5 }\OperatorTok{%>%}
\StringTok{  }\KeywordTok{mutate}\NormalTok{(}\DataTypeTok{garganta =} \KeywordTok{case_when}\NormalTok{(GARGANTA }\OperatorTok{==}\StringTok{ }\DecValTok{1} \OperatorTok{~}\StringTok{ "sim"}\NormalTok{, }\CommentTok{#yes}
\NormalTok{                              GARGANTA }\OperatorTok{==}\StringTok{ }\DecValTok{2} \OperatorTok{~}\StringTok{ "não"}\NormalTok{, }\CommentTok{#no}
                              \OtherTok{TRUE} \OperatorTok{~}\StringTok{ }\OtherTok{NA_character_}\NormalTok{))}

\CommentTok{#Dyspnea}
\NormalTok{dados5 <-}\StringTok{  }\NormalTok{dados5 }\OperatorTok{%>%}
\StringTok{  }\KeywordTok{mutate}\NormalTok{(}\DataTypeTok{dispneia =} \KeywordTok{case_when}\NormalTok{(DISPNEIA }\OperatorTok{==}\StringTok{ }\DecValTok{1} \OperatorTok{~}\StringTok{ "sim"}\NormalTok{, }\CommentTok{#yes}
\NormalTok{                              DISPNEIA }\OperatorTok{==}\StringTok{ }\DecValTok{2} \OperatorTok{~}\StringTok{ "não"}\NormalTok{, }\CommentTok{#no}
                              \OtherTok{TRUE} \OperatorTok{~}\StringTok{ }\OtherTok{NA_character_}\NormalTok{))}

\CommentTok{#Respiratory distress}
\NormalTok{dados5 <-}\StringTok{  }\NormalTok{dados5 }\OperatorTok{%>%}
\StringTok{  }\KeywordTok{mutate}\NormalTok{(}\DataTypeTok{desc_resp =} \KeywordTok{case_when}\NormalTok{(DESC_RESP }\OperatorTok{==}\StringTok{ }\DecValTok{1} \OperatorTok{~}\StringTok{ "sim"}\NormalTok{, }\CommentTok{#yes}
\NormalTok{                               DESC_RESP }\OperatorTok{==}\StringTok{ }\DecValTok{2} \OperatorTok{~}\StringTok{ "não"}\NormalTok{, }\CommentTok{#no}
                               \OtherTok{TRUE} \OperatorTok{~}\StringTok{ }\OtherTok{NA_character_}\NormalTok{))}

\CommentTok{#O2 saturation less than 95%}
\NormalTok{dados5 <-}\StringTok{  }\NormalTok{dados5 }\OperatorTok{%>%}
\StringTok{  }\KeywordTok{mutate}\NormalTok{(}\DataTypeTok{saturacao =} \KeywordTok{case_when}\NormalTok{(SATURACAO }\OperatorTok{==}\StringTok{ }\DecValTok{1} \OperatorTok{~}\StringTok{ "sim"}\NormalTok{, }\CommentTok{#yes}
\NormalTok{                               SATURACAO }\OperatorTok{==}\StringTok{ }\DecValTok{2} \OperatorTok{~}\StringTok{ "não"}\NormalTok{, }\CommentTok{#no}
                               \OtherTok{TRUE} \OperatorTok{~}\StringTok{ }\OtherTok{NA_character_}\NormalTok{))}

\CommentTok{#Diarrhea}
\NormalTok{dados5 <-}\StringTok{  }\NormalTok{dados5 }\OperatorTok{%>%}
\StringTok{  }\KeywordTok{mutate}\NormalTok{(}\DataTypeTok{diarreia =} \KeywordTok{case_when}\NormalTok{(DIARREIA }\OperatorTok{==}\StringTok{ }\DecValTok{1} \OperatorTok{~}\StringTok{ "sim"}\NormalTok{, }\CommentTok{#yes}
\NormalTok{                              DIARREIA }\OperatorTok{==}\StringTok{ }\DecValTok{2} \OperatorTok{~}\StringTok{ "não"}\NormalTok{, }\CommentTok{#no}
                              \OtherTok{TRUE} \OperatorTok{~}\StringTok{ }\OtherTok{NA_character_}\NormalTok{))}

\CommentTok{#Vomiting}
\NormalTok{dados5 <-}\StringTok{  }\NormalTok{dados5 }\OperatorTok{%>%}
\StringTok{  }\KeywordTok{mutate}\NormalTok{(}\DataTypeTok{vomito =} \KeywordTok{case_when}\NormalTok{(VOMITO }\OperatorTok{==}\StringTok{ }\DecValTok{1} \OperatorTok{~}\StringTok{ "sim"}\NormalTok{, }\CommentTok{#yes}
\NormalTok{                            VOMITO }\OperatorTok{==}\StringTok{ }\DecValTok{2} \OperatorTok{~}\StringTok{ "não"}\NormalTok{, }\CommentTok{#no}
                            \OtherTok{TRUE} \OperatorTok{~}\StringTok{ }\OtherTok{NA_character_}\NormalTok{))}

\CommentTok{#Abdominal pain}
\NormalTok{dados5 <-}\StringTok{  }\NormalTok{dados5 }\OperatorTok{%>%}
\StringTok{  }\KeywordTok{mutate}\NormalTok{(}\DataTypeTok{dor_abd =} \KeywordTok{case_when}\NormalTok{(DOR_ABD }\OperatorTok{==}\StringTok{ }\DecValTok{1} \OperatorTok{~}\StringTok{ "sim"}\NormalTok{,}
\NormalTok{                             DOR_ABD }\OperatorTok{==}\StringTok{ }\DecValTok{2} \OperatorTok{~}\StringTok{ "não"}\NormalTok{,}
                             \OtherTok{TRUE} \OperatorTok{~}\StringTok{ }\OtherTok{NA_character_}\NormalTok{))}

\CommentTok{#Fatigue}
\NormalTok{dados5 <-}\StringTok{  }\NormalTok{dados5 }\OperatorTok{%>%}
\StringTok{  }\KeywordTok{mutate}\NormalTok{(}\DataTypeTok{fadiga =} \KeywordTok{case_when}\NormalTok{(FADIGA }\OperatorTok{==}\StringTok{ }\DecValTok{1} \OperatorTok{~}\StringTok{ "sim"}\NormalTok{, }\CommentTok{#yes}
\NormalTok{                            FADIGA }\OperatorTok{==}\StringTok{ }\DecValTok{2} \OperatorTok{~}\StringTok{ "não"}\NormalTok{, }\CommentTok{#no}
                            \OtherTok{TRUE} \OperatorTok{~}\StringTok{ }\OtherTok{NA_character_}\NormalTok{))}

\CommentTok{#Olfactory loss}
\NormalTok{dados5 <-}\StringTok{  }\NormalTok{dados5 }\OperatorTok{%>%}
\StringTok{  }\KeywordTok{mutate}\NormalTok{(}\DataTypeTok{perd_olft =} \KeywordTok{case_when}\NormalTok{(PERD_OLFT }\OperatorTok{==}\StringTok{ }\DecValTok{1} \OperatorTok{~}\StringTok{ "sim"}\NormalTok{, }\CommentTok{#yes}
\NormalTok{                               PERD_OLFT }\OperatorTok{==}\StringTok{ }\DecValTok{2} \OperatorTok{~}\StringTok{ "não"}\NormalTok{, }\CommentTok{#no}
                               \OtherTok{TRUE} \OperatorTok{~}\StringTok{ }\OtherTok{NA_character_}\NormalTok{))}

\CommentTok{#Loss of taste}
\NormalTok{dados5 <-}\StringTok{  }\NormalTok{dados5 }\OperatorTok{%>%}
\StringTok{  }\KeywordTok{mutate}\NormalTok{(}\DataTypeTok{perd_pala =} \KeywordTok{case_when}\NormalTok{(PERD_PALA }\OperatorTok{==}\StringTok{ }\DecValTok{1} \OperatorTok{~}\StringTok{ "sim"}\NormalTok{, }\CommentTok{#yes}
\NormalTok{                               PERD_PALA }\OperatorTok{==}\StringTok{ }\DecValTok{2} \OperatorTok{~}\StringTok{ "não"}\NormalTok{, }\CommentTok{#no}
                               \OtherTok{TRUE} \OperatorTok{~}\StringTok{ }\OtherTok{NA_character_}\NormalTok{))}
\end{Highlighting}
\end{Shaded}

The processing of comorbidity variables is presented below.

\begin{Shaded}
\begin{Highlighting}[]
\CommentTok{#Cardiovascular disease}
\NormalTok{dados5 <-}\StringTok{  }\NormalTok{dados5 }\OperatorTok{%>%}
\StringTok{  }\KeywordTok{mutate}\NormalTok{(}\DataTypeTok{cardiopati =} \KeywordTok{case_when}\NormalTok{(CARDIOPATI }\OperatorTok{==}\StringTok{ }\DecValTok{1} \OperatorTok{~}\StringTok{ "sim"}\NormalTok{, }\CommentTok{#yes}
\NormalTok{                                CARDIOPATI }\OperatorTok{==}\StringTok{ }\DecValTok{2} \OperatorTok{~}\StringTok{ "não"}\NormalTok{, }\CommentTok{#no}
                                \OtherTok{TRUE} \OperatorTok{~}\StringTok{ }\OtherTok{NA_character_}\NormalTok{))}

\CommentTok{#Hematological disease}
\NormalTok{dados5 <-}\StringTok{  }\NormalTok{dados5 }\OperatorTok{%>%}
\StringTok{  }\KeywordTok{mutate}\NormalTok{(}\DataTypeTok{hematologi =} \KeywordTok{case_when}\NormalTok{(HEMATOLOGI }\OperatorTok{==}\StringTok{ }\DecValTok{1} \OperatorTok{~}\StringTok{ "sim"}\NormalTok{, }\CommentTok{#yes}
\NormalTok{                                HEMATOLOGI }\OperatorTok{==}\StringTok{ }\DecValTok{2} \OperatorTok{~}\StringTok{ "não"}\NormalTok{, }\CommentTok{#no}
                                \OtherTok{TRUE} \OperatorTok{~}\StringTok{ }\OtherTok{NA_character_}\NormalTok{))}

\CommentTok{#Liver disease}
\NormalTok{dados5 <-}\StringTok{  }\NormalTok{dados5 }\OperatorTok{%>%}
\StringTok{  }\KeywordTok{mutate}\NormalTok{(}\DataTypeTok{hepatica =} \KeywordTok{case_when}\NormalTok{(HEPATICA }\OperatorTok{==}\StringTok{ }\DecValTok{1} \OperatorTok{~}\StringTok{ "sim"}\NormalTok{, }\CommentTok{#yes}
\NormalTok{                              HEPATICA }\OperatorTok{==}\StringTok{ }\DecValTok{2} \OperatorTok{~}\StringTok{ "não"}\NormalTok{, }\CommentTok{#no}
                              \OtherTok{TRUE} \OperatorTok{~}\StringTok{ }\OtherTok{NA_character_}\NormalTok{))}

\CommentTok{#Asthma}
\NormalTok{dados5 <-}\StringTok{  }\NormalTok{dados5 }\OperatorTok{%>%}
\StringTok{  }\KeywordTok{mutate}\NormalTok{(}\DataTypeTok{asma =} \KeywordTok{case_when}\NormalTok{(ASMA }\OperatorTok{==}\StringTok{ }\DecValTok{1} \OperatorTok{~}\StringTok{ "sim"}\NormalTok{, }\CommentTok{#yes}
\NormalTok{                          ASMA }\OperatorTok{==}\StringTok{ }\DecValTok{2} \OperatorTok{~}\StringTok{ "não"}\NormalTok{, }\CommentTok{#no}
                          \OtherTok{TRUE} \OperatorTok{~}\StringTok{ }\OtherTok{NA_character_}\NormalTok{))}

\CommentTok{#Diabetes}
\NormalTok{dados5 <-}\StringTok{  }\NormalTok{dados5 }\OperatorTok{%>%}
\StringTok{  }\KeywordTok{mutate}\NormalTok{(}\DataTypeTok{diabetes =} \KeywordTok{case_when}\NormalTok{(DIABETES }\OperatorTok{==}\StringTok{ }\DecValTok{1} \OperatorTok{~}\StringTok{ "sim"}\NormalTok{, }\CommentTok{#yes}
\NormalTok{                              DIABETES }\OperatorTok{==}\StringTok{ }\DecValTok{2} \OperatorTok{~}\StringTok{ "não"}\NormalTok{, }\CommentTok{#no}
                              \OtherTok{TRUE} \OperatorTok{~}\StringTok{ }\OtherTok{NA_character_}\NormalTok{))}

\CommentTok{#Neurological disease}
\NormalTok{dados5 <-}\StringTok{  }\NormalTok{dados5 }\OperatorTok{%>%}
\StringTok{  }\KeywordTok{mutate}\NormalTok{(}\DataTypeTok{neuro =} \KeywordTok{case_when}\NormalTok{(NEUROLOGIC }\OperatorTok{==}\StringTok{ }\DecValTok{1} \OperatorTok{~}\StringTok{ "sim"}\NormalTok{, }\CommentTok{#yes}
\NormalTok{                           NEUROLOGIC }\OperatorTok{==}\StringTok{ }\DecValTok{2} \OperatorTok{~}\StringTok{ "não"}\NormalTok{, }\CommentTok{#no}
                           \OtherTok{TRUE} \OperatorTok{~}\StringTok{ }\OtherTok{NA_character_}\NormalTok{))}

\CommentTok{#Pneumopathy}
\NormalTok{dados5 <-}\StringTok{  }\NormalTok{dados5 }\OperatorTok{%>%}
\StringTok{  }\KeywordTok{mutate}\NormalTok{(}\DataTypeTok{pneumopati =} \KeywordTok{case_when}\NormalTok{(PNEUMOPATI }\OperatorTok{==}\StringTok{ }\DecValTok{1} \OperatorTok{~}\StringTok{ "sim"}\NormalTok{, }\CommentTok{#yes}
\NormalTok{                                PNEUMOPATI }\OperatorTok{==}\StringTok{ }\DecValTok{2} \OperatorTok{~}\StringTok{ "não"}\NormalTok{, }\CommentTok{#no}
                                \OtherTok{TRUE} \OperatorTok{~}\StringTok{ }\OtherTok{NA_character_}\NormalTok{))}

\CommentTok{#Immunosuppression}
\NormalTok{dados5 <-}\StringTok{  }\NormalTok{dados5 }\OperatorTok{%>%}
\StringTok{  }\KeywordTok{mutate}\NormalTok{(}\DataTypeTok{imunodepre =} \KeywordTok{case_when}\NormalTok{(IMUNODEPRE }\OperatorTok{==}\StringTok{ }\DecValTok{1} \OperatorTok{~}\StringTok{ "sim"}\NormalTok{, }\CommentTok{#yes}
\NormalTok{                                IMUNODEPRE }\OperatorTok{==}\StringTok{ }\DecValTok{2} \OperatorTok{~}\StringTok{ "não"}\NormalTok{, }\CommentTok{#no}
                                \OtherTok{TRUE} \OperatorTok{~}\StringTok{ }\OtherTok{NA_character_}\NormalTok{))}

\CommentTok{#Kidney disease}
\NormalTok{dados5 <-}\StringTok{  }\NormalTok{dados5 }\OperatorTok{%>%}
\StringTok{  }\KeywordTok{mutate}\NormalTok{(}\DataTypeTok{renal =} \KeywordTok{case_when}\NormalTok{(RENAL }\OperatorTok{==}\StringTok{ }\DecValTok{1} \OperatorTok{~}\StringTok{ "sim"}\NormalTok{, }\CommentTok{#yes}
\NormalTok{                           RENAL }\OperatorTok{==}\StringTok{ }\DecValTok{2} \OperatorTok{~}\StringTok{ "não"}\NormalTok{, }\CommentTok{#no}
                           \OtherTok{TRUE} \OperatorTok{~}\StringTok{ }\OtherTok{NA_character_}\NormalTok{))}

\CommentTok{#Obesity}
\NormalTok{dados5 <-}\StringTok{  }\NormalTok{dados5 }\OperatorTok{%>%}
\StringTok{  }\KeywordTok{mutate}\NormalTok{(}\DataTypeTok{obesidade =} \KeywordTok{case_when}\NormalTok{(OBESIDADE }\OperatorTok{==}\StringTok{ }\DecValTok{1} \OperatorTok{~}\StringTok{ "sim"}\NormalTok{, }\CommentTok{#yes}
\NormalTok{                               OBESIDADE }\OperatorTok{==}\StringTok{ }\DecValTok{2} \OperatorTok{~}\StringTok{ "não"}\NormalTok{, }\CommentTok{#no}
                               \OtherTok{TRUE} \OperatorTok{~}\StringTok{ }\OtherTok{NA_character_}\NormalTok{))}
\end{Highlighting}
\end{Shaded}

The processing of the variables admission to the ICU, use of ventilatory
support and evolution (death or cure) is done as follows

\begin{Shaded}
\begin{Highlighting}[]
\CommentTok{#ICU}
\NormalTok{dados5 <-}\StringTok{ }\NormalTok{dados5 }\OperatorTok{%>%}
\StringTok{  }\KeywordTok{mutate}\NormalTok{(}\DataTypeTok{uti =} \KeywordTok{case_when}\NormalTok{(UTI }\OperatorTok{==}\StringTok{ }\DecValTok{1} \OperatorTok{~}\StringTok{ "sim"}\NormalTok{, }\CommentTok{#yes}
\NormalTok{                         UTI }\OperatorTok{==}\StringTok{ }\DecValTok{2} \OperatorTok{~}\StringTok{ "não"}\NormalTok{, }\CommentTok{#no}
                         \OtherTok{TRUE} \OperatorTok{~}\StringTok{ }\OtherTok{NA_character_}\NormalTok{))}

\CommentTok{#Use of ventilatory support}
\NormalTok{dados5 <-}\StringTok{ }\NormalTok{dados5 }\OperatorTok{%>%}
\StringTok{  }\KeywordTok{mutate}\NormalTok{(}
    \DataTypeTok{suport_ven =} \KeywordTok{case_when}\NormalTok{(}
\NormalTok{      SUPORT_VEN }\OperatorTok{==}\StringTok{ }\DecValTok{1} \OperatorTok{~}\StringTok{ "invasivo"}\NormalTok{, }\CommentTok{#invasive}
\NormalTok{      SUPORT_VEN }\OperatorTok{==}\StringTok{ }\DecValTok{2} \OperatorTok{~}\StringTok{ "não invasivo"}\NormalTok{, }\CommentTok{#non-invasive}
\NormalTok{      SUPORT_VEN }\OperatorTok{==}\StringTok{ }\DecValTok{3} \OperatorTok{~}\StringTok{ "não"}\NormalTok{, }\CommentTok{#no}
      \OtherTok{TRUE} \OperatorTok{~}\StringTok{ }\OtherTok{NA_character_}
\NormalTok{    )}
\NormalTok{  )}

\NormalTok{dados5}\OperatorTok{$}\NormalTok{suport_ven <-}\StringTok{ }\KeywordTok{factor}\NormalTok{(dados5}\OperatorTok{$}\NormalTok{suport_ven,}
                            \DataTypeTok{levels =} \KeywordTok{c}\NormalTok{(}\StringTok{"invasivo"}\NormalTok{, }\StringTok{"não invasivo"}\NormalTok{, }\StringTok{"não"}\NormalTok{))}

\CommentTok{#Evolution}
\NormalTok{dados5 <-}
\StringTok{  }\NormalTok{dados5 }\OperatorTok{%>%}\StringTok{ }\KeywordTok{mutate}\NormalTok{(}
    \DataTypeTok{evolucao =} \KeywordTok{case_when}\NormalTok{(}
\NormalTok{      EVOLUCAO }\OperatorTok{==}\StringTok{ }\DecValTok{1} \OperatorTok{~}\StringTok{ "Cura"}\NormalTok{, }\CommentTok{#cure}
\NormalTok{      EVOLUCAO }\OperatorTok{==}\StringTok{ }\DecValTok{2} \OperatorTok{~}\StringTok{ "Obito"}\NormalTok{, }\CommentTok{#death }
\NormalTok{      EVOLUCAO }\OperatorTok{==}\StringTok{ }\DecValTok{3} \OperatorTok{~}\StringTok{ "Obito"}\NormalTok{, }\CommentTok{#death}
      \OtherTok{TRUE} \OperatorTok{~}\StringTok{ }\OtherTok{NA_character_}
\NormalTok{    )}
\NormalTok{  )}
\end{Highlighting}
\end{Shaded}

The analyzes obtained after the processes described above are presented
in
\url{https://observatorioobstetrico.shinyapps.io/covid_gesta_puerp_br}.

\hypertarget{final-remarks}{%
\section{4. Final remarks}\label{final-remarks}}

In this article we present the documentation for loading the data, merge
data from the years 2020 and 2021, selection and filtering cases and
processing the variables to obtain the analyzes in OOBr COVID-19,
available in
\url{https://observatorioobstetrico.shinyapps.io/covid_gesta_puerp_br}.

With OOBr COVID-19 we hope that information about COVID-19 in the
Brazilian maternal population will be accessible so that society is
aware of the pandemic situation in the country and that public policy
decisions are based on reliable data.

\hypertarget{funding}{%
\section{Funding}\label{funding}}

OOBr is financed by the Bill \& Melinda Gates Foundation, CNPq (National
Council for Scientific and Technological Development), DECIT/MS
(Department of Science and Technology of the Ministry of Health of
Brazil) and FAPES (Espírito Santo Research and Innovation Support
Foundation).

\hypertarget{references}{%
\section{References}\label{references}}

{[}1{]} Zambrano LD, Ellington S, Strid P, Galang RR, Oduyebo T, Tong
VT, Woodworth KR, Nahabedian JF 3rd, Azziz-Baumgartner E, Gilboa SM,
Meaney-Delman D; CDC COVID-19 Response Pregnancy and Infant Linked
Outcomes Team. Update: Characteristics of Symptomatic Women of
Reproductive Age with Laboratory-Confirmed SARS-CoV-2 Infection by
Pregnancy Status - United States, January 22-October 3, 2020. MMWR Morb
Mortal Wkly Rep.~2020 Nov 6;69(44):1641-1647. doi:
10.15585/mmwr.mm6944e3. PMID: 33151921; PMCID: PMC7643892

{[}2{]} Brasil. Ministério da Saúde. Departamento de informática.
{[}Open data SUS System{]} {[}Internet{]}.

\url{https://opendatasus.saude.gov.br/dataset/bd-srag-2020}, accessed
April 26, 2021.

{[}3{]} Brasil. Ministério da Saúde. Departamento de informática.
{[}Open data SUS System{]} {[}Internet{]}.

\url{https://opendatasus.saude.gov.br/dataset/bd-srag-2021}, accessed
April 26, 2021.

\end{document}